\documentclass[aps,prl,showpacs,nofootinbib,floatfix,twocolumn]{revtex4}
  
\usepackage{bm}
\usepackage{latexsym}
\usepackage{dcolumn}
\usepackage{amsfonts,amssymb,amsmath}
\usepackage{graphicx,epsfig}
\usepackage{psfrag}
\usepackage{subfigure}

\def\beq{\begin{equation}}
\def\eeq{\end{equation}}
\def\br{\begin{eqnarray}}
\def\er{\end{eqnarray}}
\def\benu{\begin{enumerate}}
\def\eenu{\end{enumerate}}

\def\apj{ApJ}
\def\mnras{MNRAS}
\def\jcap{JCAP}
\def\araa{Annu.\ Rev.\ Astron.\ Astrophys.}
\def\prd{Phys.\ Rev.\  D}
\def\physrep{Phys.\ Rep.}

\def\vk{{\bf k}}
\def\vka{{\bf k}_{1}}
\def\vkb{{\bf k}_{2}}
\def\vkc{{\bf k}_{3}}
\def\ka{k_{1}}
\def\kb{k_{2}}
\def\kc{k_{3}}

\def\cB{{\cal B}}
\def\fnl{f_{_{\rm NL}}}

\def\F{\mathcal{F}}

\begin{document}
\title{Primordial Non-Gaussianity in the Forest: 3D Bispectrum of Ly-$\alpha$ Flux Spectra Along Multiple Lines of Sight}
\author{Dhiraj Kumar Hazra\footnote{E-mail:~dhiraj@hri.res.in},
Tapomoy Guha Sarkar\footnote{E-mail:~tapomoy@hri.res.in}}
\affiliation{Harish-Chandra Research Institute, Chhatnag Road,
Jhunsi, Allahabad~211019, India.}
\date{\today}
\begin{abstract}
We investigate the possibility of constraining primordial
non-Gaussianity using the 3D bispectrum of Ly-$\alpha$ forest.  The
strength of the quadratic non-Gaussian correction to an otherwise
Gaussian primordial gravitational field is assumed to be dictated by a
single parameter $\fnl$. We present the first prediction for bounds on
$\fnl$ using Ly-$\alpha$ flux spectra along multiple lines of sight.  The
3D Ly-$\alpha$ transmitted flux field is modeled as a biased tracer of
the underlying matter distribution sampled along 1D skewers
corresponding to quasars sight lines. The precision to which $\fnl$
can be constrained depends on the survey volume, pixel noise and
aliasing noise (arising from discrete sampling of the density
field). We consider various combinations of these factors to predict
bounds on $\fnl$.  We find that in an idealized situation of full sky
survey and negligible Poisson noise one may constrain $\fnl \sim 23$
in the equilateral limit.  Assuming a Ly-$\alpha$ survey covering
large parts of the sky ($k_{min} = 8 \times 10^{-4} {\rm Mpc}^{-1}$)
and with a quasar density of ${\bar n} = 5 \times 10^{-3} \rm Mpc
^{-2}$ it is possible to constrain $ \fnl \sim 100$ for equilateral
configurations. The possibility of measuring $\fnl$ at a precision
comparable to LSS studies maybe useful for joint constraining of
inflationary scenarios using different data sets.

\end{abstract}
\pacs{98.80.Cq,  98.62.Ra}
\maketitle
{\it Introduction:}
The widely popular paradigm of slow-roll inflation driven by a single
canonical scalar field generates adiabatic perturbations which are
largely Gaussian in nature and leads to a nearly scale invariant power
spectrum ~\cite{maldacena-2003}. Several theoretical predictions
however point towards mild to severe departure from Gaussianity
\cite{bartolo-2004}. Measuring the degree of non-Gaussianity is hence
crucial towards discriminating between various inflationary scenarios
thereby enhancing our understanding of the very early Universe.  It is
assumed that on sub-horizon scales the primordial gravitational
potential $\Phi^{\rm prim}$ is related to a Gaussian random field
$\phi_G$ through a non linear relation of the form $ \Phi^{\rm prim} =
\phi_G + \frac{\fnl}{c^2} \left( {\phi_G}^2 -\langle {\phi_{G}}^2
\rangle \right)$.  Departures from Gaussianity may hence be quantified
using the parameter $\fnl$.  We assume that $\fnl$ is scale
independent - a reasonable prediction for most inflationary models
where non-Gaussianity is generated on super-horizon scales. The value
of $\fnl$ obtained from slow-roll inflation turns out to be very small
$({\cal O} ~( 10^{-3} - 10^{-2} ))$ \cite{maldacena-2003}. This
implies that any detection of large $\fnl$ shall rule out all
canonical single field slow-roll inflation models.  The mean value of
$\fnl$ ($26\pm140$ for equilateral and $32\pm21$ in local limit for
$1\sigma$ CL) obtained from WMAP data~\cite{wmap-7} seems to indicate
large non-Gaussianity. Although low SNR in these results indicate that
we are yet to detect the primordial non-Gaussianity, it is expected
that data from the Planck satellite\footnote{http://www.sciops.esa.int/PLANCK/}
shall give a much tighter constraint on $\fnl$ and the error is
expected to come down to $\Delta \fnl \sim \pm5$ in the local
limit. Other than CMBR observations, a measurement of the bispectrum
or the three point correlation function of the galaxy distribution
\cite{bslss} is a standard alternative method to constrain primordial
non-Gaussianity. These probes however only provide weak bounds on
$\fnl$ as compared to the CMBR observations
(SDSS\footnote{http://www.sdss.org/} for example can measure $|\fnl|
\sim 10^3 - 10^4$).

In the post reionization epoch, small fluctuations of the neutral
hydrogen (HI) density field in a predominantly ionized IGM leads to a
series of distinct absorption features - the Ly-$\alpha$ forest
\cite{rauch1998} in the spectra of background quasars. The Ly-$\alpha$
forest is a well established powerful probe of cosmology
\cite{lyalphacosm, psbs}.  Traditional Ly-$\alpha$ studies have
considered the power spectrum or bispectrum of the one dimensional
transmitted flux field \cite{psbs} corresponding to the quasar line of
sight. This approach is reasonable when the angular density of quasars
on the sky is low. The new generation of quasar surveys (the ongoing
BOSS\footnote{http://cosmology.lbl.gov/BOSS} and future
BigBOSS\footnote{http://bigboss.lbl.gov/index.html}) however promise
to achieve a very high quasar density and cover large fractions of the
sky. This has led to the possibility of measuring the 3D Ly-$\alpha$
power spectrum along multiple lines of sight \cite{multiplelos}.
 We note that the first hydro simulation of the Ly-$\alpha$ forest in
non-Gaussian scenarios is presented in an earlier work \cite{viel09}.

In this {\it letter} we investigate the possibility of constraining $\fnl$
using the 3D Ly-$\alpha$ forest bispectrum. Similar to the power
spectrum studies,  the Ly-$\alpha$ flux distribution is assumed to be a
biased tracer of the underlying matter field sampled along discrete
sight lines.  We explore the range of observational parameters for
the constraints on $\fnl$ from the 3D analysis to be competitive with
CMBR and LSS (large scale structure) studies.

{\it Formalism:}
The post-reionization matter density field $\Delta_{\vk}$ in Fourier
space is related to the primordial gravitational potential on
sub-horizon scales as $ \Delta_{\vk} (z) = {\cal{M}} (k, z) \Phi^{\rm
  prim}_{\vk}$. The function $ {\cal M} (k, z)$ is given by $
{\cal{M}}(k, z) = -\frac{3}{5} \frac{k^2 T(k)}{\Omega_m H_0^2}
D_{+}(z)$, where $T(k)$ denotes the matter transfer function and
$D_{+}(z)$ is the growing mode of density fluctuations.  We have used
the BBKS transfer function~\cite{bbks} and cosmological
parameters~\cite{hazra-2010} obtained from a
MCMC \footnote{http://cosmologist.info/cosmomc/} analysis on the WMAP7
data-sets. The power spectrum and bispectrum of the density field are
defined as $ \langle \Delta_{\vka}\Delta_{\vkb} \rangle =
\delta_D(\vka + \vkb) P(\ka)$ and $ \langle \Delta_{\vka}
\Delta_{\vkb} \Delta_{\vkc}\rangle = \delta_D(\vka + \vkb + \vkc)
B(\ka, \kb, \kc)$.  Clearly, the linear power spectrum of the density
field is given by $P^{\rm L} (k)= {\cal{M}} (k, z)^2 P_{\Phi}^{\rm
  prim}$ where $ P_{\Phi}^{\rm prim}$ denotes the primordial power
spectrum of the gravitational potential such that $ P_{\Phi}^{\rm
  prim} = P_{\phi {G}} + {\cal{O}} (\fnl ^2)$.  The powerspectrum $
P_{\phi {G}} $ of the Gaussian field $\phi_G$ shall be assumed to be
featureless and scale invariant. We note that a wide class of inflation models,
including the simplest one comprising of a single
inflaton field in a quadratic potential, introduces perturbations that
are almost Gaussian and exhibit a power spectrum that is nearly scale
invariant. It follows that in the realm of linear perturbation theory
the bispectrum of the matter field arising from primordial
non-Gaussianity is given by ${B^{\rm L}}_{123} =
{\cal{M}}(\ka){\cal{M}}(\kb){\cal{M}}(\kc) {B_{\phi G}}_{123} $ where we
use the notation ${123} \equiv (\ka, \kb, \kc)$ and $ B_{\phi G}$ is
given by \br {B_{\phi G}}_{123}= \frac{2 \fnl}{c^2}\left [ P_{\phi G}(\ka)
  P_{\phi G} (\kb) + \rm {cyc}\right] + {\cal{O}} (\fnl ^3).  \er Apart
from the contribution to the bispectrum from primordial fluctuations,
non-linear structure formation caused by gravitational instability
leads to mode coupling and thereby induce additional non-Gaussianity.
This is especially relevant when we use low redshift tracers to
implicitly measure $n-$point functions of the matter density field.
Using the second order perturbation theory the additional contribution
to the matter bispectrum is \beq B^{\rm NL}_{123} = 2 F_2(\vka, \vkb)
P(\ka) P(\kb) + \rm{cyc}.  \eeq where the second order correction is
obtained from perturbation theory and we have adopted the form of
$F_2$ from \cite{bslss}.  Finally, the total matter bispectrum is a
sum of the contributions to non-Gaussianity arising from the intrinsic
primordial fluctuations and that generated by non-linear evolution of
an otherwise Gaussian field. Thus we have $B_{123} = B^{\rm L}_{123} +
B^{\rm NL}_{123}$, where we have ignored the possible contribution
from the primordial trispectrum.  We use $B_{123}$ to obtain the 3D
bispectrum of the Ly-$\alpha$ forest.

The Ly-$\alpha$ forest spectra are associated with gas distribution in
voids or slightly overdense regions.  Noting that the astrophysical
structures associated with the spectra are only mildly non-linear, the
transmitted flux $\F$ through the Ly-$\alpha$ forest may be modeled by
assuming that the gas traces the underlying dark matter distribution
\cite{psbs} except on small scales where pressure plays an important
role.  Further, it is believed that photo-ionization equilibrium that
maintains the neutral fraction also leads to a power law
temperature-density relation \cite{tempdens}. The fluctuating
Gunn-Peterson approximation (FGPA)\cite{fgpa} incorporates these
assumptions to relate the transmitted flux $\F$ to the dark matter
overdensity $\delta$ as $ {\F} = {\bar{\F}} ~{\rm exp}\left [- A ( 1 +
  \delta ) ^{ 2 - 0.7(\gamma -1)} \right ]$, where ${\bar{\F}}$ is the
mean transmitted flux and $(\gamma -1)$ is the slope of the
temperature-density relation \cite{tempdens}.  We note that $\gamma$
imprints the reionization history of the Universe. The redshift
dependent quantity $A$ \cite{bolton} depends on a number of parameters,
like the IGM temperature, photo-ionization rate and cosmological
parameters \cite {psbs}.  On large scales it is reasonable to believe
that the fluctuations in the transmitted flux $\delta_{\F} = \left (
\F/ \bar{\F} - 1 \right )$ may be expanded as $ \delta_{\F} = b_1
\delta + \frac{1}{2} b_2 \delta^2 $ where, it is assumed that the
Ly-$\alpha$ forest spectrum has been smoothed over some suitably large
length scale. This relation allows analytic computation of statistical
properties of $\delta_{\F}$.  We note that corrections to this on
small scales come from peculiar velocities, an effect we have not
incorporated in our analysis for simplicity.  At our fiducial redshift
$ z = 2.5$, we adopt an approximate $(\bar{\F}, \gamma, A) \equiv (0.8, \, 1.5, \, 0.16)$ from the numerical simulations of Ly-$\alpha$ forest
\cite{mcd03} and theoretical predictions \cite{seljak12}. We note
however that these numbers are largely uncertain owing to inadequate
modeling of the IGM. The bias, $b_1$ for example has a sensitive
redshift dependence and may depend on the smoothing scale of the
Ly-$\alpha$ spectra.  Using the local bias model, the power spectrum
$P_{\F}(k)$ and bispectrum ${\cal{B}}_{\F}$ of Ly-$\alpha$ forest flux
fluctuations $\delta_{\F}$ are given by \br P_{\F}(k) &=& b_1^2 P(k)
\nonumber \\ \cB_{\F 123} &=& b_1^3 B_{123} + b_1^2 b_2 \left [P(\ka)
  P(\kb) + \rm{cyc}. \right ] \er The bispectrum of Ly-$\alpha$ flux
is hence completely modeled using three parameters $(\fnl, b_1, b_2)$.
We shall now set up the Fisher matrix for constraining $\fnl$ using
the Ly-$\alpha$ bispectrum.  Following the formulation described in
\cite{bslss} we define the bispectrum estimator as \br \hat{\cB}_{\F
  123} = \frac{V_f}{V_{123}} \int_{\ka} d^3 {\bf q}_1 \int_{\kb} d^3
     {\bf q}_2 \int_{\kc} d^3 {\bf q}_3 \delta_D( {\bf q}_{123})
     \nonumber \\ \times
     ~\Delta_{\F}^o(\ka)\Delta_{\F}^o(\kb)\Delta_{\F}^{o} (\kc) \er
     Here $ {\bf q}_{123} = {\bf q}_{1} + {\bf q}_{2} + {\bf{ q}}_{3}
     $, ${V_f} = (2 \pi)^3/V$ where $V$ is the survey volume, $V_{123}
     = \int_{\ka} d^3 {\bf q}_1 \int_{\kb} d^3 {\bf q}_2 \int_{\kc}
     d^3 {\bf q}_3 \delta_D( {\bf q}_{123})$ and the integrals are
     performed over the $q_i-$ intervals $\left (k_i - \frac{\delta
       k}{2} , k_i + \frac{\delta k}{2}\right )$. The quantities
     $\Delta_{\F}^o(k_i)$ denotes the `observed' Ly-$\alpha$ flux
     fluctuations in Fourier space.  The observed quantity
     $\delta_{{\F}}^o({\bf{r}})$ is given by the continuous field
     $\delta_{{\F}}({\bf r})$ sampled along skewers corresponding to
     line of sight to bright quasars. We therefore have
     $\delta_{{\F}}^o({\bf{r}})=\delta_{{\F} }({\bf{r}}) \times
     \rho({\bf{r}}) $, where the sampling window function
     $\rho({\bf{r}})$ is defined as $\rho({\bf{r}}) = {\cal{N}}
     \frac{\sum_a w_a \ \delta_D^{2}( {\bf{r}}_{\perp} -
       {\bf{r}}_{\perp a}) }{\sum_a w_a}$, where ${\cal{N}}$ is a
     normalization such that $\int dV \rho({\bf{r}}) = 1$.  The
     summation extends up to $N_{Q}$, the total number of quasar
     skewers in the field which are assumed to be distributed with sky locations
     ${\bf{r}}_{\perp a}$ . The weights $w_a$ introduced in
     $\rho({\bf{r}})$ are in general related to the pixel noise and
     can be chosen with a posteriori criterion of minimizing the
     variance.  In Fourier space, we then have $
     {\Delta}_{{\F}}^o({\bf{k}}) = \tilde{\rho}({\bf{k}}) \otimes
     {\Delta_{\F}}({\bf{k}}) + \Delta_{\F \rm noise}({\bf{k}})$, where
     $\tilde{\rho}$ is the Fourier transform of $\rho$, and
     $\Delta_{\F \rm noise}({\bf{k}})$ denotes a possible noise term.

 If the bispectrum covariance matrix is diagonal which implies that no correlation exists between different triangle shapes,  
the simple  variance of the estimator $\hat{\cB_{\F}}$ can be calculated as $\Delta\hat{\cB_{\F}}^2 = \langle \hat{\cB_{\F}}^2 \rangle - \langle \hat{\cB_{\F}} \rangle ^2$.
This is given at the lowest order  by
\beq
\Delta\hat{\cB _{\F}}^2 = \frac{V_f}{V_{123}} s  P_{\F}^{\rm Tot}(\ka) P_{\F}^{\rm Tot}(\kb) P_{\F}^{\rm Tot}(\kc)
\eeq
where $s = 6, 1$ for equilateral and scalene triangles respectively and  $P_{\F}^{\rm Tot}(k)$ is the total power spectrum of Ly-$\alpha$ flux given by
\beq
 P_{\F}^{\rm Tot}({\bf k} ) = P_{\F}({\bf k}) + P^{\rm {1D}}_{\F}(k_{\parallel}) P_{W} + N_{\F}
\label{eq:totalps}
\eeq The quantity $P^{\rm 1D}_{\F}(k_\parallel)$ is the usual 1D flux
power spectrum \cite{psbs} corresponding to individual spectra given
by $P^{\rm {1D}}_{\F}(k_{\parallel}) = (2\pi)^{-2} \int d^2{\bf
  k}_{\perp} P_{\F}({\bf k})$ and $ P_{W} $ denotes the
power spectrum of the window function. The quantity $ N^{}_{\F}$
denotes the effective noise power spectra for the Ly-$\alpha$
observations.  The term $ P^{\rm {1D}}_{\F}(k_{\parallel}) P_{W}$ referred to as the `aliasing' term, is similar to the shot
noise in galaxy surveys and quantifies the discreteness of the 1D
Ly-$\alpha$ skewers.  It has been shown that  an uniform weighing scheme
suffices when most of the spectra are measured with a sufficiently
high SNR \cite{multiplelos}. This gives $ P_{W} =
\frac{1}{\bar{n}}$, where $\bar{n}$ is the 2D density of quasars
($\bar{n} = N_{Q}/ {\mathcal{A}}$, where ${\mathcal{A}}$ is the area
of the observed field of view).  We assume that the variance
$\sigma^2_{{\F} N}$ of the pixel noise contribution to $\delta_{\F}$
is the same across all the quasar spectra whereby we have $N_{\F} =
\sigma^{2}_{\F N}/\bar{n}$ for its noise power spectrum.  In arriving
at equation (\ref{eq:totalps}) we have ignored the effect of quasar
clustering. In reality, the distribution of quasars is expected to
exhibit clustering \cite{myers}. However, for the quasar surveys under
consideration, the Poisson noise dominates over the clustering and the
latter may be ignored.

The Fisher matrix for a set of parameters $p_i$ is constructed as \beq
F_{ij} = \sum_{\ka = k_{min}}^{k_{max}} \sum_{\kb = k_{min}}^{\ka}
\sum_{\kc = \tilde{k}_{min}}^{\kb} \frac{1}{\Delta\hat{\cB_{\F}}^2}
\frac{\partial{\cB_{\F 123}}}{\partial p_i}\frac{\partial{\cB_{\F
      123}}}{\partial p_j}\eeq

where $ \tilde{k}_{min} = max(k_{min}, |\ka - \kb|)$ and the summations are performed using $\delta k = k_{min}$. Assuming the likelihood function for $p_i$ to be a 
Gaussian the errors in $p_i$ is given by the Cramer-Rao bound
$\sigma_i^2 = F^{-1}_{ii}$.
We have used this to investigate the  power of a Ly-$\alpha$ survey to constrain $\fnl$.

{\it Results:}
 We consider quasars in the range  $z=2$ to $3$ since the peak in redshift distribution of quasars occur in this range \cite{sneider}.
 We note that for a given quasar at redshift $z = z_Q$, the proximity effect will not allow the spectrum to be measured in the  region 
 $10,000 \, {\rm km \, s^{-1}}$ blue-wards of the  Ly-$\alpha$
emission and only the region which is  at least 1,000 $\rm km  \, 
s^{-1}$ red-ward of the quasar's Ly-$\beta$ and O-$\rm VI$  lines 
are considered to avoid the possible confusion with these lines.
We have chosen $z = 2.5$ as out fiducial redshift for the subsequent analysis. We note here that all the parameters involved in the modeling the 
 Ly-$\alpha$ forest, have direct or indirect redshift dependence.

A Ly-$\alpha$ forest survey towards measurement of power spectrum or bispectrum is characterized by  the survey volume, 
pixel noise in the spectra and the number density of the quasar skewers. 
The constraining power of the survey shall depend directly on the choice of these parameters.
In the cosmic variance limit the minimum $\fnl$ that can be measured depends on the number of Fourier modes in the survey volume V given by $N_k = 4 \pi / 3  k_{max}^2 V/(2 \pi)^3$.
Clearly the minimum detectable $\fnl$  is a function of  $k_{max}$ and $k_{min}$.
The noise power spectrum $ N_{\F}$ is given by $ N_{\F} = {\bar \F}^{ \, -2} [S/N]^{-2}_{\Delta x} (\Delta x/ 1 {\rm Mpc})$ where $ [S/N]_{\Delta x} $ is the signal to noise ratio 
for a spectrum smoothed to a resolution $\Delta x$. We quote $[S/N]$  here for   $1~\mathring{A}$ pixels.
The main source of noise to the 3D power spectrum comes from the aliasing noise term and one requires a very high density of quasars in the field of 
view for this term to be sub-dominant. 

\noindent
\begin{figure}[!t]
\begin{center}

\resizebox{80pt}{65pt}{\includegraphics{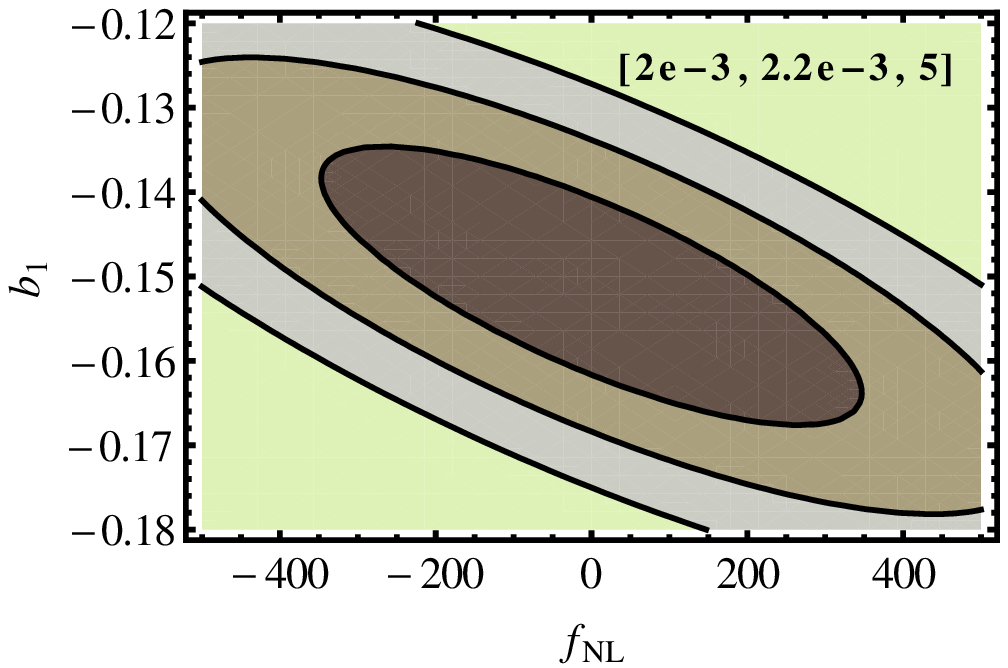}}
\resizebox{80pt}{65pt}{\includegraphics{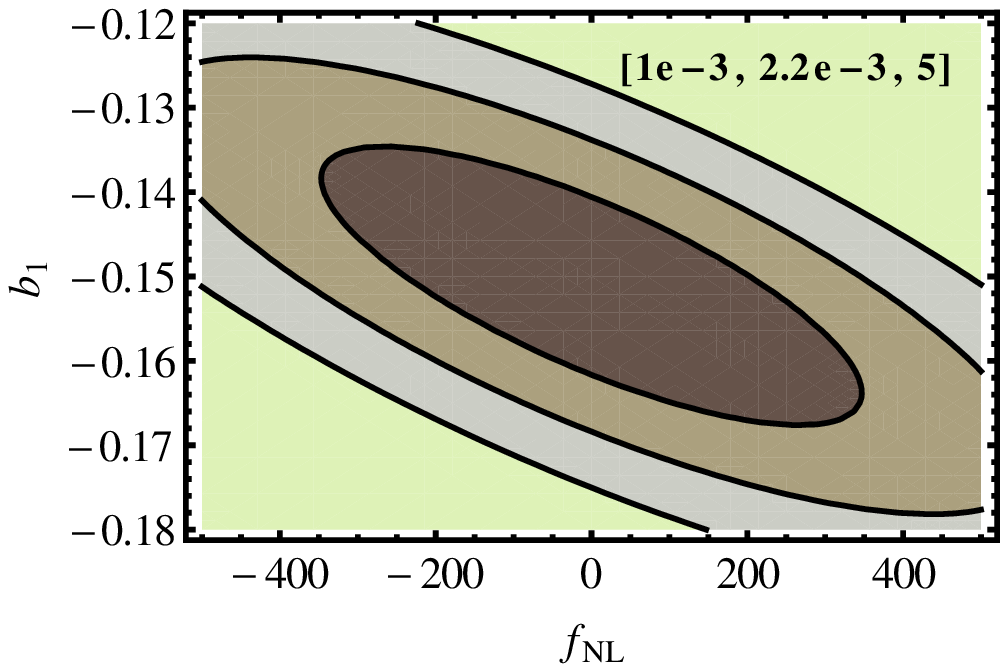}}
\resizebox{80pt}{65pt}{\includegraphics{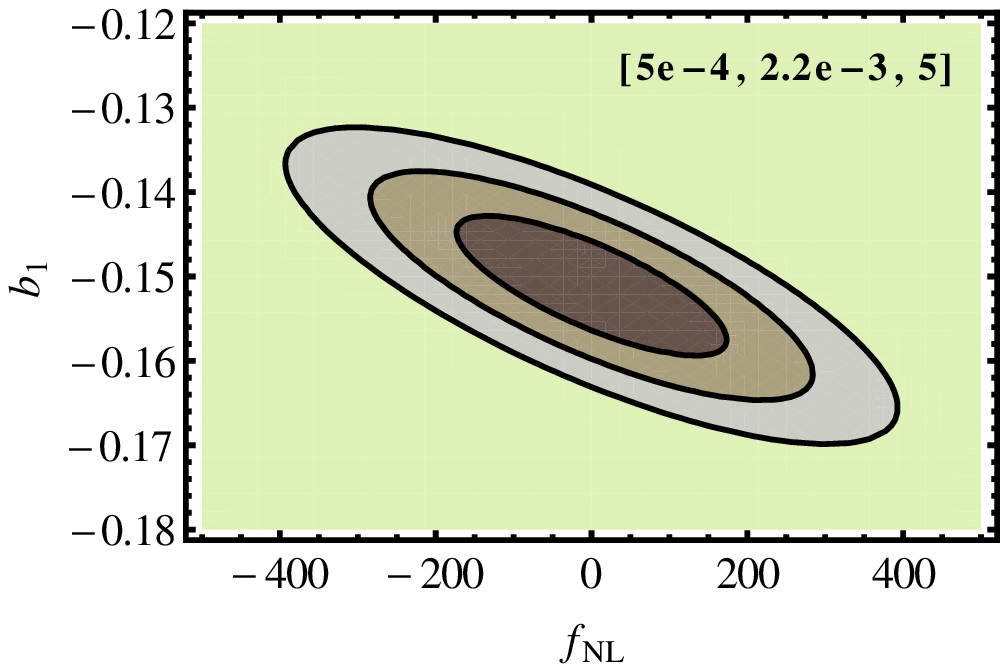}}

\resizebox{80pt}{65pt}{\includegraphics{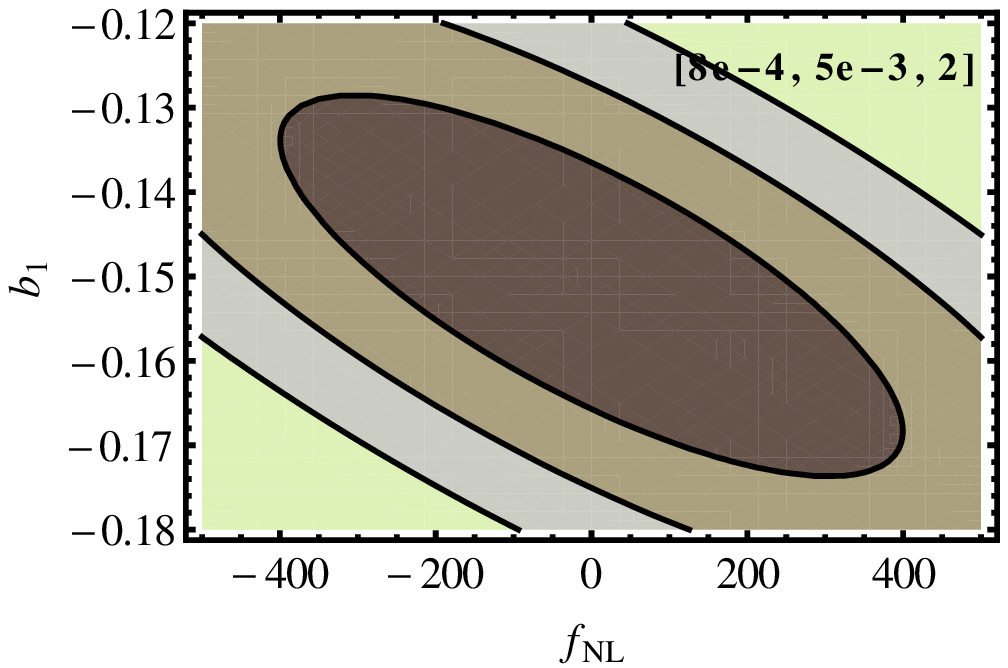}}
\resizebox{80pt}{65pt}{\includegraphics{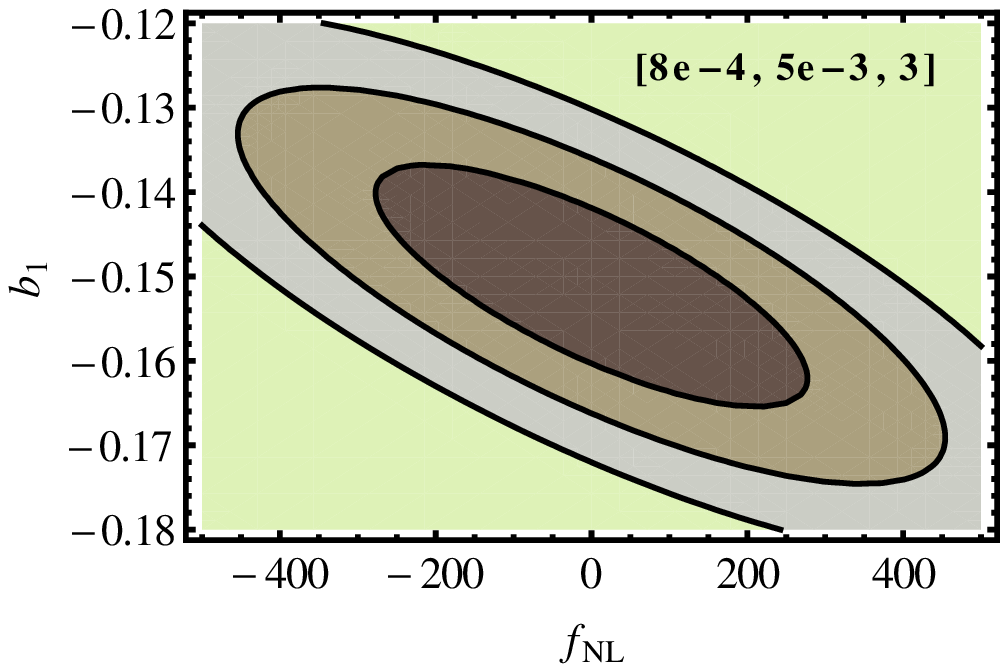}}
\resizebox{80pt}{65pt}{\includegraphics{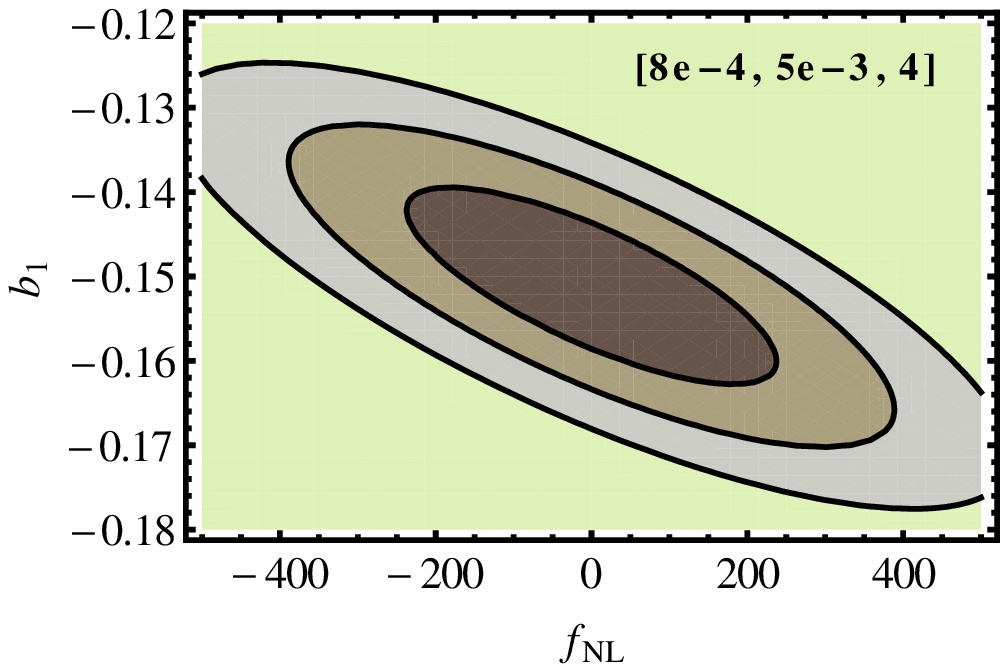}}

\resizebox{80pt}{65pt}{\includegraphics{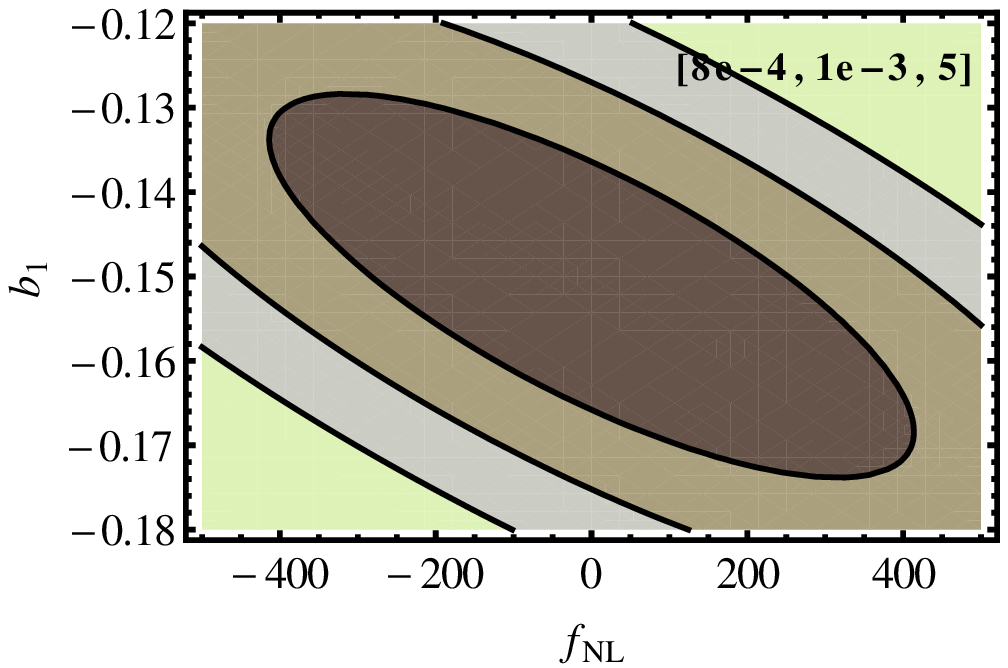}}
\resizebox{80pt}{65pt}{\includegraphics{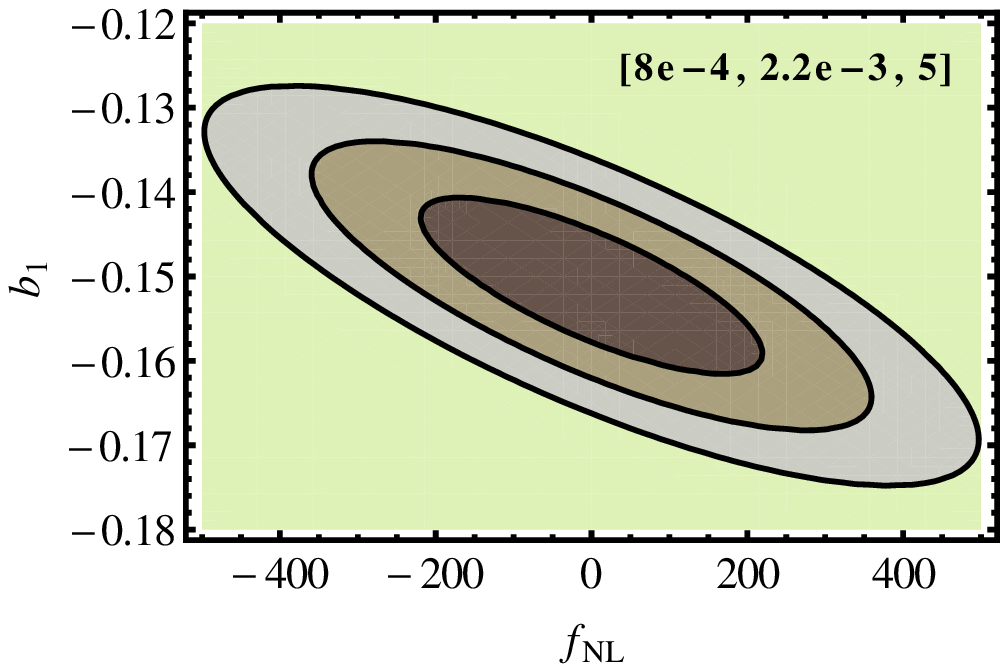}}
\resizebox{80pt}{65pt}{\includegraphics{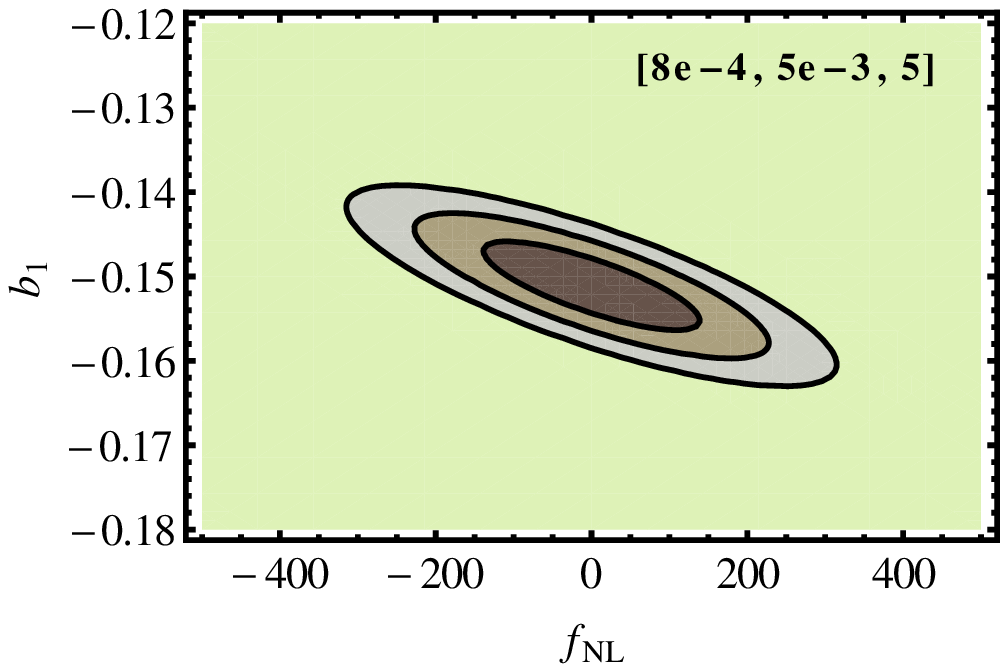}}
\end{center}
\caption{The  $68.3 \%$, $95.4\%$ and $99.8\%$ likelihood confidence contours for the parameters $(\fnl, b_1)$. Shown in the figure
  are the values $(k_{min}, {\bar n} ,  S/N)$ used to compute the Fisher matrix.}
\label{fig:contours}
\end{figure}
\begin{table}[!htb]
\begin{center}
\begin{tabular}{c c c c c}
\hline\hline
 & & & &\\
$k_{min}$  & ${\bar n}$  &$S/N$ & $\Delta\fnl$ & $\Delta b_1$ \\
 (${\rm Mpc^{-1}}$) & (${\rm Mpc^{-2}}$) &$ $ & $ $ & $ $ \\
 
 \hline\hline

 & & & &\\

$2\times10^{-3}~~~~$ & $~~~~2.2\times10^{-3}~~~~$ &$5$ &$~~~~228.84~~$ &$~~~~1.1\times10^{-2}$\\
$1\times10^{-3}~~~~$ & $~~~~2.2\times10^{-3}~~~~$ &$5$ &$~~~~161.81~~$ &$~~~~7.7\times10^{-3}$\\
$5\times10^{-4}~~~~$ & $~~~~2.2\times10^{-3}~~~~$  & $5$& $~~~~114.42~~$&$~~~~5.5\times10^{-3}$\\

\hline
$8\times10^{-4}~~~~$ & $~~~~1.0\times10^{-3}~~~~$ & $5$&$~~~~272.95~~$ &$~~~~1.5\times10^{-2} $\\
$8\times10^{-4}~~~~$ & $~~~~2.2\times10^{-3}~~~~$ & $5$ &$~~~~144.73~~$ &$~~~~6.9\times10^{-3}$ \\
$8\times10^{-4}~~~~$ & $~~~~5.0\times10^{-3}~~~~$ & $5$ &$~~~~91.65~~$ &$~~~~3.5\times10^{-3}$\\

\hline

$8\times10^{-4}~~~~$ & $~~~~2.2\times10^{-3}~~~~$ & $2$& $~~~~263.52~~$&$~~~~1.5\times10^{-2} $\\
$8\times10^{-4}~~~~$ & $~~~~2.2\times10^{-3}~~~~$ & $3$&$~~~~182.83~~$ & $~~~~9.5\times10^{-3}$\\
$8\times10^{-4}~~~~$ & $~~~~2.2\times10^{-3}~~~~$ & $4$&$~~~~156.56~~$ &$~~~~7.7\times10^{-3}$\\

\hline
{\bf \it Ideal case} & & & &\\
$5\times10^{-4}~~~~$ & $~~~~1~~~~$ & $5$&$~~~~23.72$ &$~~~~2.1\times10^{-4}$\\
\hline\hline
\end{tabular}
\end{center}
\caption{\label{tab:results}The  bounds on ($\fnl,~b_1$) obtained from Fisher analysis for various combinations of ($k_{min},~{\bar n},~S/N$).}
\end{table}
The bispectrum SNR depends on the triangle configurations considered
to evaluate it. In this work we have used the simplest equilateral
configurations characterized by just a single Fourier mode. This over
estimates the noise by at least a factor of $\sim 2.45$ as compared to
the case with arbitrary triangles.  In the equilateral limit the 3D
Ly-$\alpha$ bispectrum can be written as $ \cB_{\F}(k) = P(k)^2 \left[
  \frac{a_1}{{\cal M} (k)} + a_2\right]$ where $ a_1 = 6b_1^3 \fnl/
c^2$ and $a_2 = 6b_1^3F_2 + 3 b_1^2b_2$. Only two parameters are
sufficient to model the bispectrum instead of three parameters $(\fnl,
b_1, b_2)$ for the general case. We use the fiducial values $( \fnl ,
b_1, b_2) \equiv ( 0, -0.15, -0.075) $ and choose $\fnl$ and $b_1$ to
be the free parameters for the Fisher analysis. We recall that
  in our modeling of the Ly-$\alpha$ forest we used the parameters
  (${\bar \F}, A, \gamma$). The parameter $\bar{\F}$ does not appear in
  $\delta_{\F}$ and there is degeneracy between the parameters $A$ and $\gamma$ which only appears as a product in $b_1$. Changing $b_1$ hence amounts to
  changing either or both $A$ and $\gamma$.

We assume that the
likelihood function is a bivariate Gaussian which yields the
confidence ellipses shown in figure (\ref{fig:contours}).  The tilt of
the error ellipses indicate correlation between the parameters. We
quantify this using the correlation coefficient $ r = F^{-1}_{12}/
\sqrt{F^{-1}_{11} F^{-1}_{22}}$. For the range of parameters chosen we
find that this is roughly constant $ r \sim -0.7$.

In the ideal situation of full sky coverage and negligible Poisson
noise we find that $\Delta \fnl \sim 23$ in the equilateral limit. 
We tabulate our results for varying sky coverage ($k_{min}^{-3}=V/(2\pi)^{3}$),
Poisson noise ($\sim 1/{\bar n}$) and pixel noise ($S/N$) in
Table~\ref{tab:results}.  As expected we have tighter constraints on
($\fnl,b_1$) with increasing survey volume, ${\bar n}$ and $S/N$.  The
values of the survey parameters chosen are reasonable and achievable
by future Ly-$\alpha$ surveys. Exploiting the entire sky coverage of
SDSS we find that one can obtain a bound on $\fnl \sim 100$ (in
equilateral configuration) for a survey with ${\bar n}=5\times
10^{-3}~{\rm Mpc}^{-2}$ when the spectra are measured at $5 \sigma$
level. 
 
 Our analysis has largely focussed on the equilateral configuration.
 However we find that the Cramer-Rao bound for $f_{NL}$ in the
  squeezed limit ($ k_3 << min (k_1, k_2)$) turns out to be $\sim 40 -
  100$ for the cases we have considered.  The case of arbitrary
triangular configuration is to be addressed in our future
  work ~\cite{sarkar-prep}. However,  our preliminary estimates show us that we may
  constrain $f_{NL} \sim 1$ in an ideal environment. For example using $S/N \sim 5$,
  $\bar{n} \sim 10^{-3} \rm Mpc^{-3}$ and $k_{min} \sim 10^{-3} \rm
    Mpc^{-1}$, we have $\Delta f_{NL} \sim 5$ in the case of arbitrary triangles,
 which is competitive with CMBR and LSS studies.

   Our discussion so far has bypassed the issue of redshift space
    distortion arising from peculiar motion. The bispectrum in
    redshift space shall depend not only on the shape of the triangle
    but also on its orientation with the line of sight. This shall
    introduce two more angular parameters in the analysis. The
    multipole expansion of bispectrum in redshift space provides a way
    to break the degeneracy between bias and cosmological growth
    parameter $f \sim \Omega_m^{0.6}$. We plan to take this up in a
    future work~\cite{sarkar-prep}. However the spherically averaged bispectrum is
    related to its real space counterpart as ${\mathcal B}_{\F}^s =
    \left( 1 + \frac{2}{3}\beta + \frac{1}{9} \beta^2 \right
    ){\mathcal B}_{\F} $ where $\beta = f /b_1$ \cite{sefu}. This is
    an enhancement on all scales.

To conclude,  we emphasize that it is possible to put
stringent bounds on primordial non-Gaussianity from the measured 3D
bispectrum of the Ly-$\alpha$ forest along multiple lines of sight and
thereby constrain various inflationary scenarios. Our analytic
predictions indicate that such studies with future Ly-$\alpha$ surveys
may be useful while performing a joint analysis using other data sets
like CMBR or LSS.


\end{document}